\documentclass[aps,prb,twocolumn,superscriptaddress]{revtex4-1}
\usepackage{graphicx}
\usepackage{amsmath}
\usepackage{hyperref}

\begin{document}


\title{Study of phonons in irradiated epitaxial thin films of UO$_2$}


\author{S. Rennie}
\author{E. Lawrence Bright}
\author{J. E. Darnbrough}
\altaffiliation[Present Address: ]{University of Oxford, Materials Department, Parks Road, Oxford, OX1 3PH, UK}
\affiliation{University of Bristol Interface Analysis Centre, HH Wills Physics Laboratory, Tyndall Avenue, Bristol, BS8 1TL, UK}
\author{L.Paolasini}
\author{A. Bosak}
\affiliation{European Synchrotron Radiation Facility (ESRF), Boîte Postale 220, F-38043 Grenoble, France}
\author{A. D. Smith}
\author{N. Mason}
\affiliation{Dalton Cumbrian Facility, University of Manchester, Westlakes Science and Technology Park, Moor Row, Cumbria CA24 3HA, UK}
\author{G. H. Lander}
\affiliation{European Commission, Joint Research Centre (JRC), Directorate for Nuclear Safety and Security, Postfach 2340, D-76125 Karlsruhe, Germany}
\author{R. Springell}
\affiliation{University of Bristol Interface Analysis Centre, HH Wills Physics Laboratory, Tyndall Avenue, Bristol, BS8 1TL, UK}




\date{\today}

\begin{abstract}
We report experiments to determine the effect of radiation damage on the phonon spectra of the most common nuclear fuel, UO$_2$. We have irradiated thin ($\sim$ 300\,nm) epitaxial films of UO$_2$ with 2.1\,MeV He$^{2+}$ ions to 0.15\,dpa and a lattice swelling of $\Delta$a/a $\sim$ 0.6\,\%, and then used grazing-incidence inelastic X-ray scattering to measure the phonon spectrum. We succeeded to observe the acoustic modes, both transverse and longitudinal, across the Brillouin zone. The phonon energies, in both the pristine and irradiated samples, are unchanged from those observed in bulk material. On the other hand, the phonon linewidths (inversely proportional to the phonon lifetimes), show a significant broadening when comparing the pristine and irradiated samples. This effect is shown to increase with phonon energy across the Brillouin zone. The decreases in the phonon lifetimes of the acoustic modes are roughly consistent with a 50\% reduction in the thermal conductivity.
\end{abstract}



\maketitle

\section{Introduction}
A key limitation of using uranium dioxide as a nuclear reactor fuel is it’s intrinsically low thermal conductivity, which is known to significantly decay as a function of irradiation damage \cite{Ronchi2004,Ronchi2004a}. As the thermal conductivity falls, the radial temperature gradient across the fuel pin, becomes more substantial, leading to enhanced cracking and deformation. Consequently, the decay in thermal conductivity not only reduces the reactor efficiency but also contributes to the degradation in structural integrity of the fuel; together these effects ultimately act to limit the fuel lifetime.

The microscopic thermodynamic variables that drive the reduction in thermal conductivity have not yet been identified. Past work suggests that thermal conductivity is dominated by phonons \cite{Ronchi1999}, for temperatures below half the melting temperature (3120 K), where the quasiharmonic approximation is valid and contributions from polarons are small \cite{Ronchi1999,Harding1989}. The phonon-dispersion curves for undamaged UO$_2$ were first obtained by the pioneering measurements of Dolling \textit{et al}. in (1965) \cite{Dolling1965}. More recently, and of great significance to the present study, Pang \textit{et al}. \cite{Pang2013, Pang2014} have revisited the phonons of UO$_2$ at 295 and 1200 K and measured the phonon linewidths, as well as the phonon energies. From these measurements they extracted the thermal conductivities for each phonon branch, and showed that the totals are in excellent agreement with thermal conductivity measured by conventional techniques in UO$_2$ \cite{Ronchi2004}. This proves that, at these lower temperatures, the phonons are indeed the important transporters of heat, as expected. Furthermore, Pang \textit{et al}. showed that at room temperature the branch-specific thermal conductivities are roughly divided into four (almost equal) contributions from the transverse (TA), longitudinal (LA) acoustic, transverse (TO$_1$), and longitudinal (LO$_1$) optic modes. The strong involvement of the optic modes is unexpected, and not predicted by theory \cite{Pang2013}.

Given this good agreement, it would seem an obvious next step to perform the same experiments on an irradiated single crystal of UO$_2$ and large (up to 100 g) single crystals of UO$_2$ exist. However, irradiation of a crystal in a reactor will result in a dose rate of $>$\,100 R/hr (mostly from short-lived fission products) that, because of the danger to personnel, would be impossible to examine with any instrument at a neutron user facility. An alternative is to damage the crystal with charged particles from an accelerator, but such radiation does not penetrate into a bulk crystal more than several $\mu$m, so the damage would be inhomogeneous. We have overcome these difficulties by uniformly damaging thin epitaxial films of UO$_2$ with accelerated charged particles and then examined the phonons with inelastic X-ray scattering (IXS) in grazing incidence. There are clearly at least two significant challenges to be faced. (1) The first is to choose a suitable amount of damage so that some effect may be observed, and (2) to develop the technology of measuring phonons from thin films with sufficient precision to determine the linewidths.

\section{Experimental Details and Results}
\subsection{Production and characterisation of UO$_2$ epitaxial films}

A few examples of partially epitaxial UO$_2$ films can be found in the literature back before 2000, but the first major effort was undertaken at Los Alamos National Laboratory with a polymer assisted deposition method \cite{Scott2014}. The use of DC magnetron sputtering to produce epitaxial films was first reported by Strehle \textit{et al}. \cite{Strehle2012}, and such epitaxial films were fabricated by Bao \textit{et al}. at both Karlsruhe and Oxford/Bristol at about the same time \cite{Bao2013}. More details of the growth and characterization of these films can be found in Ref. \cite{Rennie2017}. Much thinner epitaxial films are used for so-called dissolution studies \cite{Springell2015}.

\begin{figure}
\centering
\includegraphics[height=0.25\textheight]{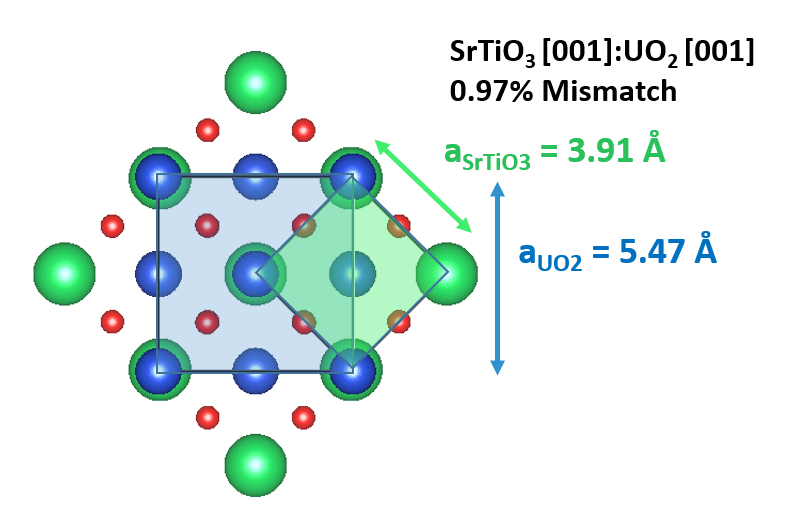}
\caption{Epitaxial (001) UO$_2$ thin films were deposited on [001] SrTiO$_3$. As depicted, the UO$_2$:SrTiO$_3$  system has a $\sqrt{2}$ epitaxial relation with a lattice mismatch of 0.97\%. Figure created using the VESTA software \cite{Momma2011}. \label{fig:fig1}}
\end{figure}

The epitaxial films of (001) UO$_2$ were produced via DC magnetron sputtering at Bristol University on substrates of (001) SrTiO$_3$ obtained commercially from MTI corp. These systems have a $\sqrt{2}$ epitaxial match, achieved through a 45$^{\circ}$, giving a lattice mismatch of 0.97$\%$, as shown in Fig. \ref{fig:fig1}. An argon pressure of 7\,$\times$\,10$^{-3}$\,mbar and an oxygen partial pressure of 2\,$\times$\,10$^{-5}$\,mbar were used to sputter epitaxial films at 1000$^{\circ}$C, giving a deposition rate of 0.2\,nm/sec.

\subsection{Radiation damage in thin epitaxial UO$_2$ films}
One of the most difficult parameters to determine was the amount and type of radiation damage to produce in the films. If the damage is too extensive and the lattice itself is partially destroyed, then clearly we are unable to measure the phonon spectra as related to crystal directions; on the other hand, too little damage risks observing only small, or no changes in the phonons.

\begin{figure*}
\centering
\includegraphics[height=0.24\textheight]{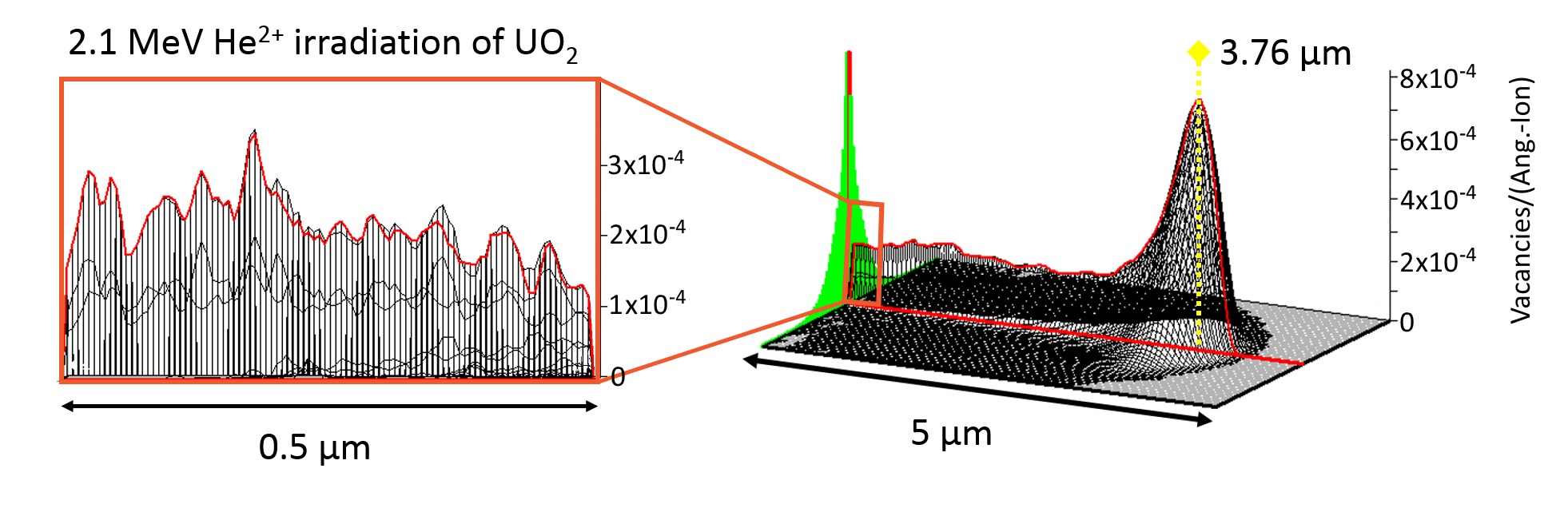}
\caption{The irradiation damage profiles calculated using the monolayer method in SRIM \cite{Ziegler2008} for the irradiation of a 0.5\,$\mu$m UO$_2$ layer and bulk UO$_2$ sample with 2.1\,MeV He$^{2+}$ ions, using displacement energies of 20 eV and 40 eV for oxygen and uranium respectively. The dashed yellow lines represent the peak of the damage, located at 3.76\,$\mu$m, i.e. in the substrate. \label{fig:fig2}}
\end{figure*}

An important aspect is the uniformity of the damage in both the growth direction and across the surface of the film, since the grazing-incidence IXS casts a sizeable footprint of several mm on the film. In Fig. \ref{fig:fig2} we show the calculated damage profile for irradiation with light ions of He$^{2+}$, and they clearly show two aspects: (1) The “Bragg peak”, i.e. the most damaged region where the high-energy ions eventually stop, is deep into the substrate, Fig \ref{fig:fig2}, and (2) over a film thickness of 500\,nm the damage distribution is homogeneous, Fig \ref{fig:fig2}, insert. During the irradiation, the ion beam was rastered, so that the entire sample was damaged uniformly in the xy plane. Light ion (He$^{2+}$) irradiations  are less likely to cause significant displacements of the heavier uranium atoms compared with the lighter oxygen due to the small momentum transfer.

Irradiation experiments were conducted at the Dalton Cumbrian Facility (DCF) using a 5 MV Tandem Pelletron Ion Accelerator \cite{Wady2015}. Samples were damaged by 2.1\,MeV He$^{2+}$ ions generated by the TORVIS and the SRIM calculated displacements per atom was 0.15\,dpa. The flux was 1.8\, $\times$\,10$^{12}$ He$^{2+}$/cm$^2$/sec, and the accumulated dose: 6.7\,$\times$\,10$^{16}$ He$^{2+}$/cm$^2$ ions.

Two identical samples were made of dimensions 5\,$\times$\,5\,mm$^2$ cross section and thickness 300\,nm. One of these (the pristine sample) was not irradiated, and throughout the study a comparison was made between the phonons deduced from the pristine and irradiated samples.

\begin{figure*}
\centering
\includegraphics[height=0.27\textheight]{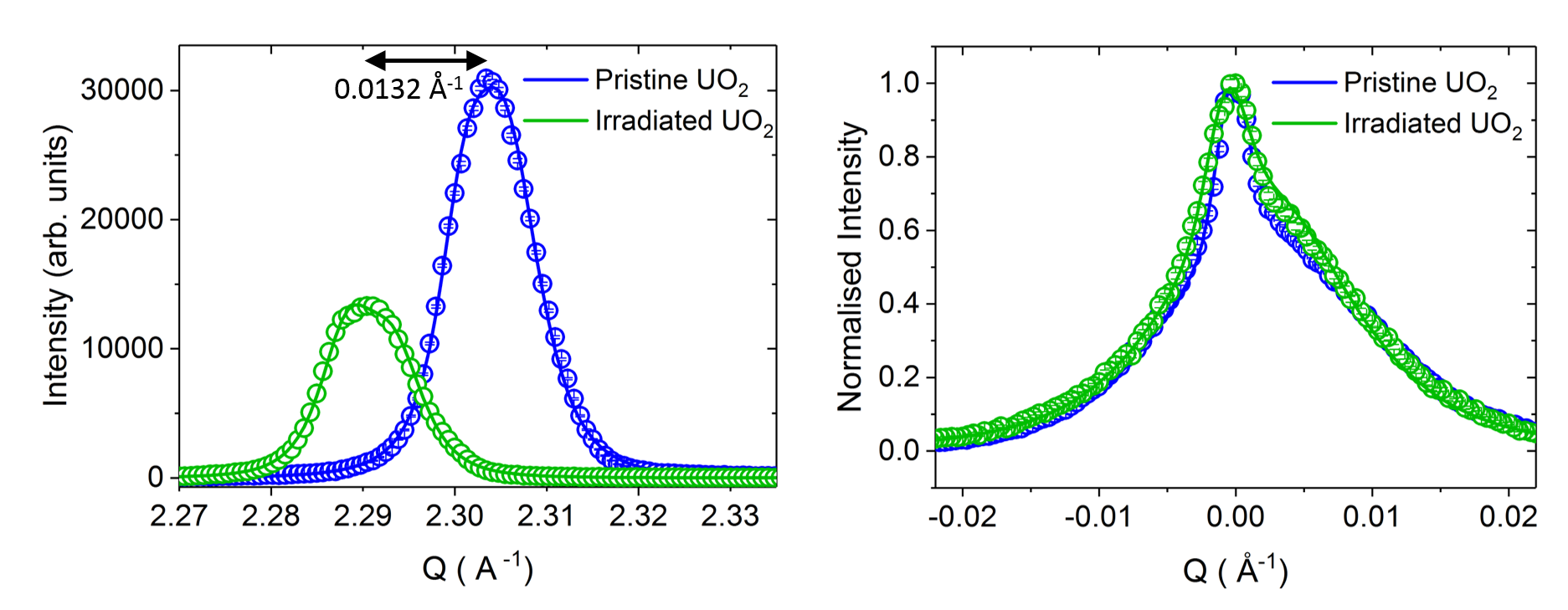}
\caption{Comparative longitudinal (i.e. $\theta$-2$\theta$) (left) and transverse ($\theta$ only) (right) diffraction profiles of the (002) reflection from the pristine (open blue circles) and irradiated films (open green circles). The shift in the longitudinal scans corresponds to a increased lattice parameter $\Delta$a/a\,=\,+\,0.56(2)\,\% for the irradiated film. There is also a small broadening of the FWHM in the transverse scans for the irradiated films.  \label{fig:fig3}}
\end{figure*}

The thin films were characterized through measurements of the (002) UO$_2$ XRD peak. These data are shown in Fig. \ref{fig:fig3}. There is a sizeable change in the lattice parameter corresponding to an expansion of $\Delta$a/a\,=\,+\,0.56(2)\,\%. Since the full-widths at half maximum (FWHM) are almost the same for the two films in both the longitudinal and transverse directions, we can conclude that the damage is uniform across the 300\,nm of the film, and the crystallinity remains almost intact. Other tests were performed on off-specular reflections, and, as expected, the UO$_2$ films were fully epitaxial.

\subsection{Measuring phonons by grazing-incidence inelastic X-ray scattering}

\begin{figure*}
\centering
\includegraphics[height=0.4\textheight]{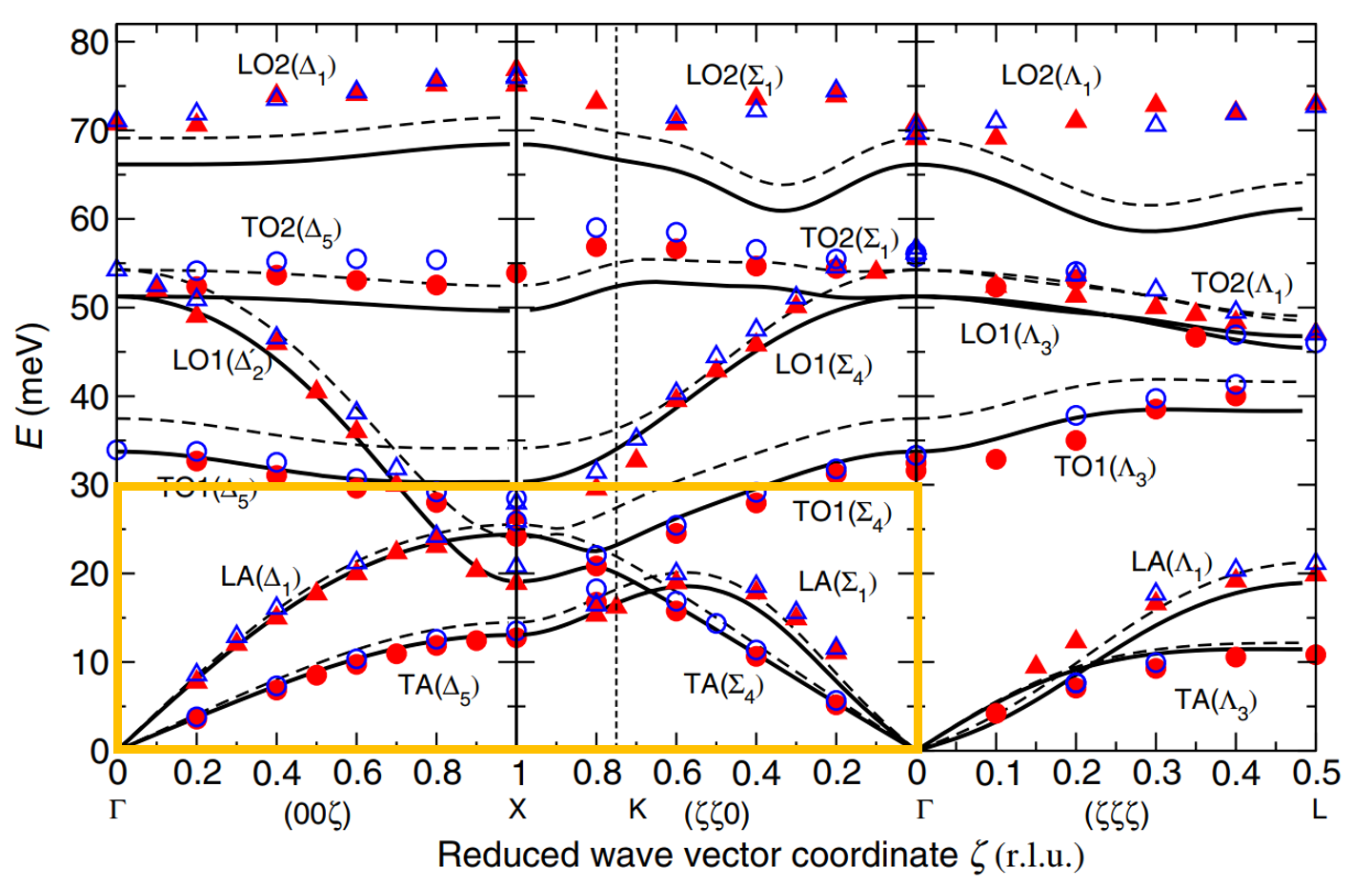}
\caption{The yellow box highlights the region of the phonon dispersion explored during the present study of thin films. Data of the full dispersion curves are taken from recent neutron work on bulk unirradiated UO$_2$, as measured by Pang \textit{et al}. \cite{Pang2013}. Measurements were taken at 295 K (blue open circles) and at 1200 K (red solid symbols), where the circles and triangles represent the transverse and longitudinal phonon modes, respectively. The solid and dashed lines are theory – see Ref. \cite{Pang2013}. \label{fig:fig4}}
\end{figure*}

The area of momentum space and energy we cover in our experiments is shown in Fig \ref{fig:fig4}, as a yellow box superimposed on the results of Ref. \cite{Pang2013} from a bulk stoichiometric sample of UO$_2$ measured by inelastic neutron scattering (INS). This region omits: (1) Any modes in the [$\zeta \zeta \zeta$] direction, and (2) any modes with energies above 30 meV. The first is related to the use of a SrTiO$_3$ substrate, which allows for the deposition of (001) oriented UO$_2$. Exploration of the [$\zeta \zeta \zeta]$ direction is possible with an (110) oriented UO$_2$ film, however while this can be grown using (110) YSZ (yttria stabilized zirconia) substrates \cite{Strehle2012}, defects within YSZ are known to give rise to significant diffuse scattering. The second is related to the challenge of seeing optic modes with IXS in any heavy metal oxide. This is demonstrated by recent work \cite{Maldonado2016} on NpO$_2$ where a small single crystal of 1.2\,mg was successfully used (in conventional reflection geometry rather than that of grazing incidence of the present work) to determine the phonons at room temperature. In this study, it was not possible to measure optic modes, as their intensity (arising mainly from oxygen displacements) is at least a factor of 100 times weaker than that from acoustic modes. Furthermore, the important mode LO$_1$, which Pang \textit{et al}.\cite{Pang2013} show carries $\sim$ 1/3 of the heat in UO$_2$, could not be observed with IXS in NpO$_2$ (See Fig. 3 in Ref \cite{Maldonado2016} and surrounding discussion); this mode is known to have no contribution from the metal atoms \cite{Elliott1967}.

The experiments to measure the phonons from the thin films at room temperature were performed on the ID28 spectrometer at the European Synchrotron Radiation Facility \cite{ESRF}. Grazing-incidence IXS was conducted with a Kirkpatrick-Baez mirror, together with a Be focusing lens to produce a focused beam of 15\,$\mu$m\,(vertical)\,$\times$\,30\,$\mu$m\,(horizontal) with an inclination angle of 0.2\,$^{\circ}$ out of the horizontal plane. The incident energy was 17.794\,keV, with an instrumental resolution of 3\,meV. This energy, is determined by reflections from the Si(999) reflections in the analyzers, is just above the \textit{U\,L$_3$} resonant energy of 17.166\,keV. This increases the absorption of the incident beam, and the 1/e penetration of the photon beam of this energy in UO$_2$ is 10.4\,$\mu$m. The vertical spot size is 15\,$\mu$m, implying at an incident angle of 0.2\,$^{\circ}$ a footprint of $\sim$\,2.5 mm. Much of the beam intensity is lost to absorption. The critical angle of UO$_2$ for this energy X-ray beam is 0.18\,$^{\circ}$,  reducing the interaction further. Given the absorption, the penetration depth of the X-ray beam will be $\sim$\,15\,nm at this angle.

\begin{figure*}
\centering
\includegraphics[height=0.65\textheight]{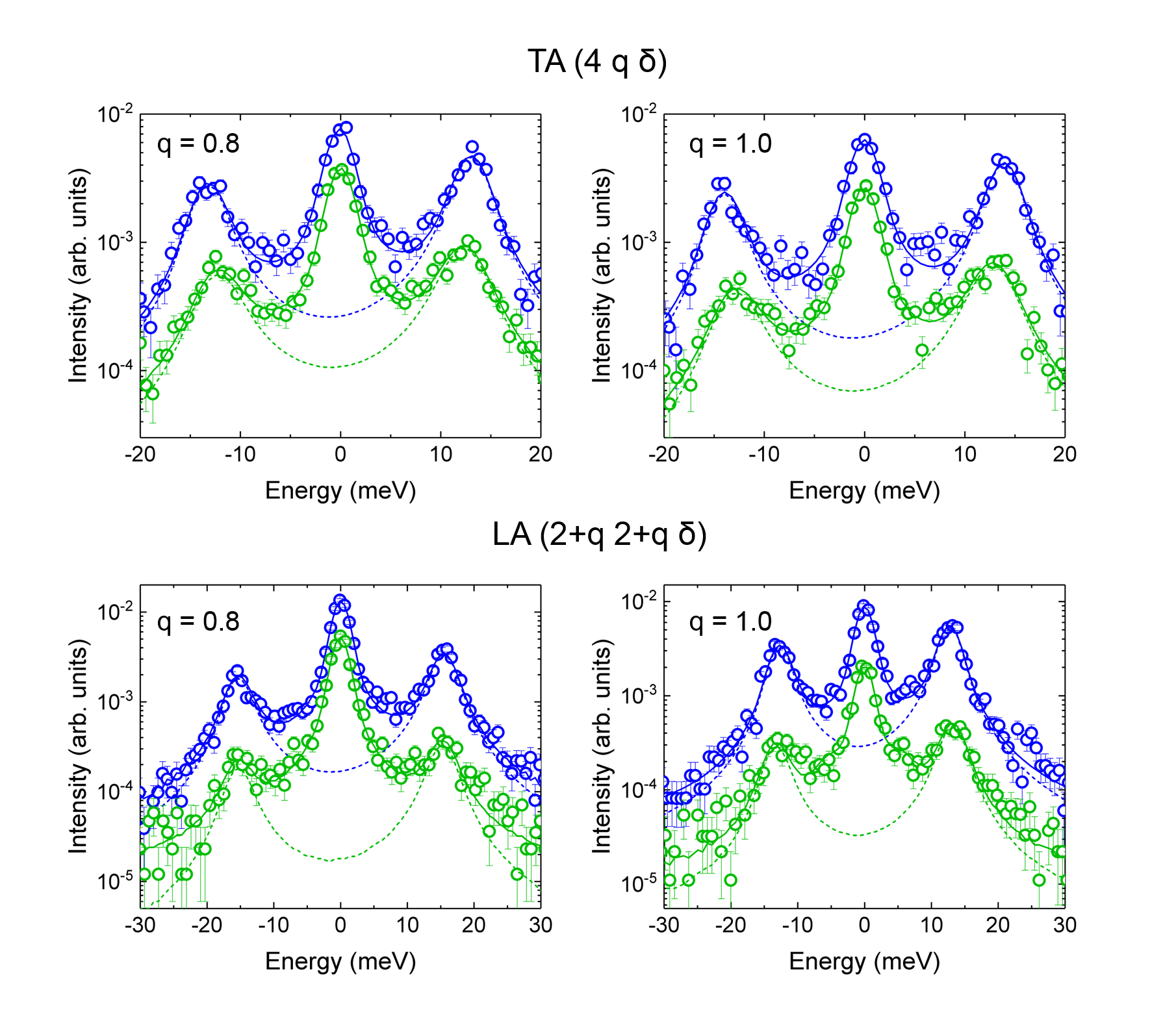}
\caption{We show (upper) phonons measured from the TA (100) at positions (4\,0.8\,$\delta$) and (4\,1.0\,$\delta$) and (lower) those from the LA (110) at positions (2.8\,2.8\,$\delta$) and (3.0\,3.0\,$\delta$). In each case the blue are pristine, and the green are irradiated samples. The fits have used a Gaussian resolution of 3\,meV convoluted with Lorentzian consisting of a central (resolution limited) peak together with a Damped Harmonic Oscillator (DHO) representing both the Stokes and anti-Stokes phonons, weighted by the Bose factors, to reproduce the experimental curves. The width of the DHO function is then giving the experimentally deduced phonon linewidths. The data have been normalized differently so that they do not overlap in the figures; the higher intensity data are from the pristine film. \label{fig:fig5}}
\end{figure*}

Two experimental efforts were made on ID28. In the first, a 450\,nm film was used, and in the second a 300\,nm film. To increase the strength of the signal in the second attempt the film was tilted an additional 0.5\,$^{\circ}$ around an axis in the horizontal plane. This results in a small L component in the observed phonon modes, i.e. not completely in the horizontal plane (HK0), where the L component is indicated by $\delta$, where 0.03\,$<$\,$\delta$\,$<$\,0.15. However, this small L component allows a deeper penetration of $\sim$\,150\,nm, and a concomitant order of magnitude increase in the phonon signals. With the small penetration depths no evidence for the substrate was seen. In Fig. \ref{fig:fig5} we show a selection of data.

Figure \ref{fig:fig5} shows both the measured Stokes and anti-Stokes phonons and this gives a better determination of the absolute energy of the excitation. The central (elastic) line arises from thermal diffuse scattering and defects and it is noticeable that it is stronger (compared to the phonons) in the irradiated samples (green) than in the pristine samples (blue), as would be anticipated.

\begin{figure*}
\centering
\includegraphics[height=0.35\textheight]{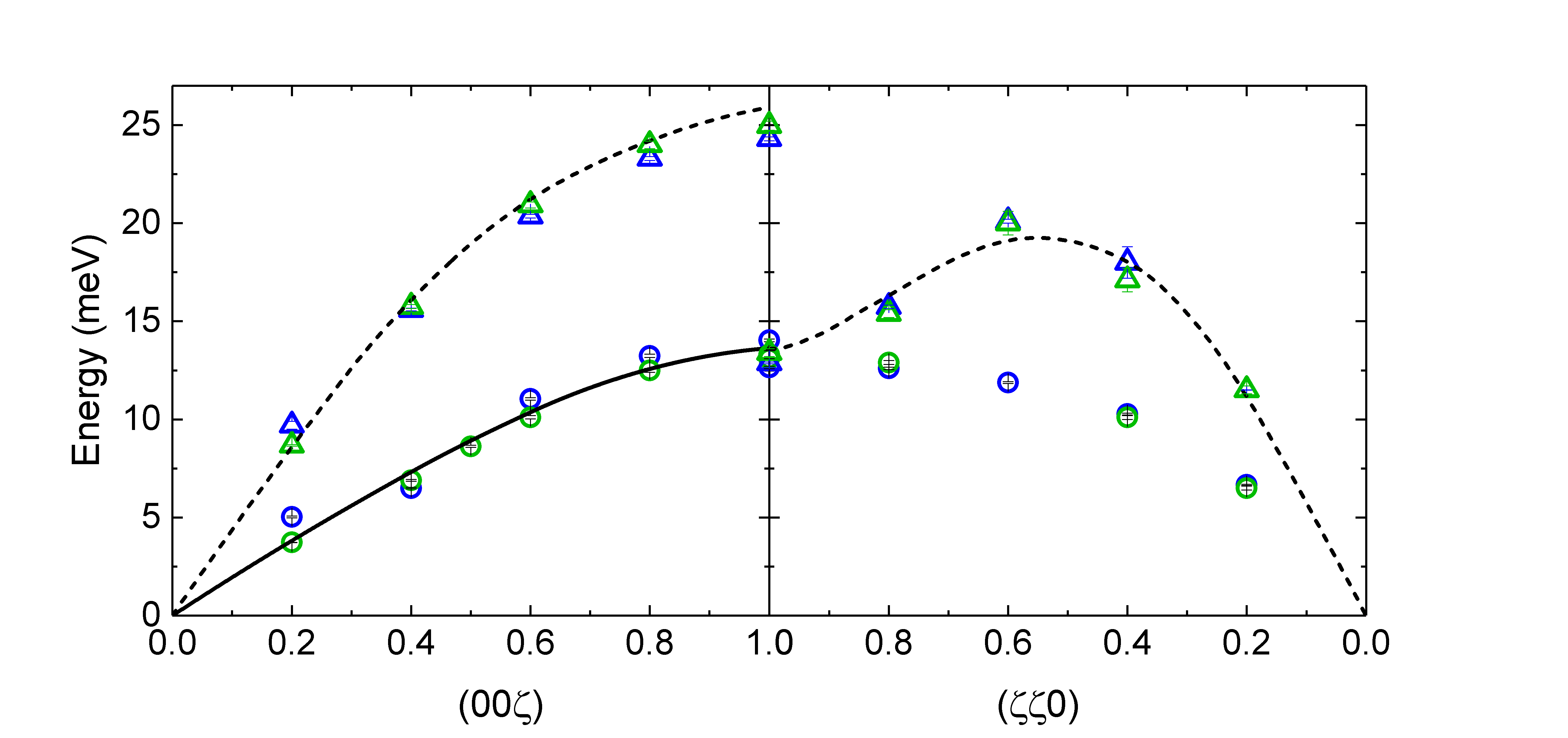}
\caption{The energies of the transverse (open circles) and longitudinal (open triangles) phonons of UO$_2$ along the (00$\zeta$) and ($\zeta\zeta$0) directions as measured via IXS for pristine (blue) and irradiated (green) thin films, in comparison with bulk data from Ref. \cite{Pang2013} where theses data (obtained by INS) have been fitted to give smooth curves represented by the solid (TA) and dashed (LA) lines. The TA and LA phonon energies at the X-point should be degenerate; the observed differences in the thin-film data are a consequence of the small L component introduced by tilting the film out of the horizontal plane. In the ($\zeta\zeta$0) direction a different TA phonon was measured compared to the INS work, due to different orientations; see text. \label{fig:fig6}}
\end{figure*}

From these analyses we determine the energy and linewidth of the phonons. Figure \ref{fig:fig6} shows the phonon energies that we have measured with the thin films at 295\,K and compares them to those reported by Pang\,\textit{et\,al}. \cite{Pang2013}.

A comparison between the data shown in Figs.\,\ref{fig:fig4} and \ref{fig:fig6} shows that in the ($\zeta\zeta$0) direction a different TA phonon has been measured in the films compared with the bulk spectra shown in Fig.\,\ref{fig:fig4}. This is because the standard orientation, as used to provide Fig.\,\ref{fig:fig4}\,\cite{Pang2013, Pang2014}, is with [1$\bar{1}$0] vertical, and our film (because of the epitaxy with the SrTiO3 substrate) has [001] vertical. The TA (110) phonons in Fig.\,\ref{fig:fig4} have a polarization [00u], whereas in our case this mode cannot be observed, and we have a TA (110) phonon with polarization [u, -u, 0] (where u is the small atomic displacement from the equilibrium atomic position). In our case all the atomic displacements in the measured phonons lie in the plane of the film. This is because in grazing incidence the scattering vector \textbf{Q} lies very close to the plane (within 2\,$^{\circ}$), and phonons are observed when the product \textbf{Q\,$\cdot$\,u} is non-zero. This makes the measurements insensitive to atomic vibrations along the film growth axis. The LA modes are the same in both our work and that of Pang\,\textit{et\,al}. \cite{Pang2013, Pang2014}.

Figure\,\ref{fig:fig6} shows no significant differences in the phonon energies between the pristine and the irradiated films, and both results agree within statistics with the energies determined by Pang\,\textit{et\,al}. \cite{Pang2013}, measured by INS.

The lack of any difference in the phonon energies between the pristine and irradiated films is not surprising, since the lattice parameter has been changed by only 0.6\,\% and the crystallinity of the sample is still preserved. Based on the well-known thermal expansion of unirradiated UO$_2$ \cite{Martin1987} the lattice expands by 0.85\,\% between 295 and 1200\,K, and we can see in Fig. \ref{fig:fig4} that this expansion makes little difference to the energies of the acoustic modes.

However, the more likely change would come in the linewidths; an increase of the linewidths would translate to a decrease in the phonon lifetimes and a concomitant decrease of thermal conductivity \cite{Pang2013, Pang2014}.

\subsection{Analysis of the phonon linewidths}
\begin{figure*}
\centering
\includegraphics[height=0.38\textheight]{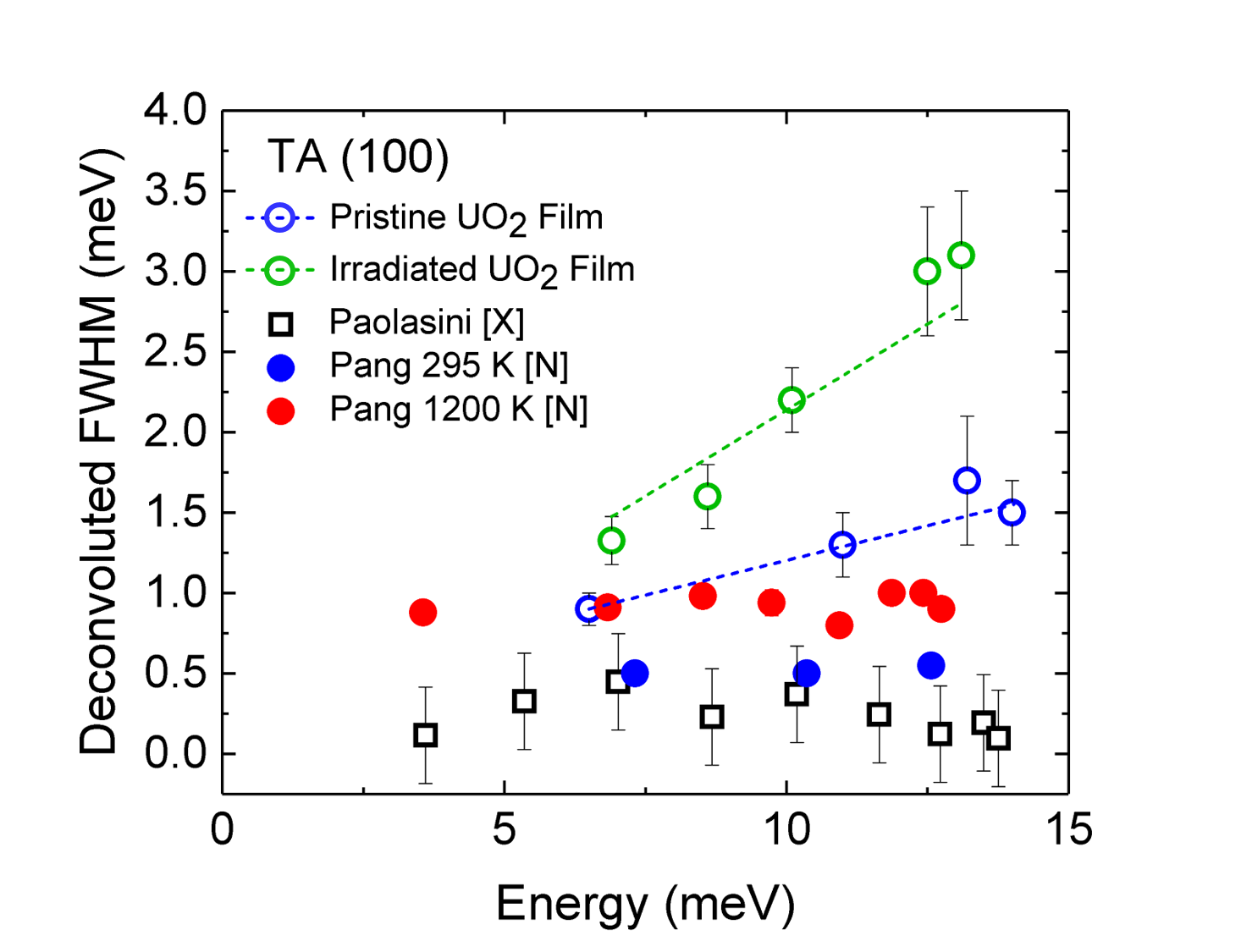}
\caption{Values of the FWHM deduced from analysis of the phonons measured in the TA [100] direction. The values tabulated in Ref. \cite{Pang2014} by INS are shown as blue (295\,K) and red (1200\,K) filled circles. Our values using IXS are shown as open blue (pristine) and open green (irradiated) circles. Values determined from a small bulk UO$_2$ crystal at room temperature determined on the same X-ray instrument are shown as open black squares \cite{Paolasini}.  \label{fig:fig7}}
\end{figure*}

The effects on phonon lifetime are present in the FWHM and are related directly (by their inverse values) to the thermal conductivity. It is important in this respect to compare our values with those deduced from bulk UO$_2$ from Tables A1 and A2 of Ref. \cite{Pang2014}. We tabulate all our measured energies and deduced linewidths in the Appendix. For graphical representation we show only the TA (100) mode – see Fig.\,\ref{fig:fig7}. The TA(110) mode is not the same as measured by Pang\,\textit{et\,al}. \cite{Pang2013,Pang2014}, and the LA modes are weaker than the TA. This reduces the statistics for the LA modes, however similar trends are seen in all acoustic modes. Fig.\,\ref{fig:fig5} shows the FWHM of low-energy phonons are nontrivial to fit as there is an appreciable contribution to their intensity from the central elastic line. Therefore the FWHM’s for the lower-energy phonons ($<$\,$\sim$\,5\,meV) are omitted from Fig.\,\ref{fig:fig7}.

\section{Discussion}

\subsection{Extent of radiation damage}
Previous work into radiation damage is broadly separated into two different aspects: damage in reactor and damage as a product of self-irradiation as a spent fuel.  The first relates to damage of the fuel, and the formation of the “high burn-up structure” in nuclear fuels \cite{Rondinella2010}. The second is long-term storage of irradiated nuclear fuel, and what happens to the fuel as a function of time \cite{Wiss2014}. In the storage case (Fig. 1 of Ref \cite{Wiss2014}) damage of $\sim$\,0.15\,dpa corresponds approximately to the activity of moderately radiated 60\,GWd/t (gigawatt – days per ton) fuel when it is removed from the reactor. This would correspond to $\sim$\,5\,$\times$\,10$^{17}$ alpha-decay events/g, and have a swelling of 0.7\,\% (Table 1 Ref.  \cite{Wiss2014}) compared with the swelling we produced of $\sim$\,0.6\% (i.e.\,$\Delta$a/a\,=\,0.56\,\%). The thermal conductivity of this material would have decreased also by $\sim$\,50\% \cite{Lucuta1996}. The precise relationship between the lattice swelling, the concentration of defects, and the drop in thermal conductivity is still open for discussion \cite{Staicu2010}.

With alpha particles, He$^{2+}$ ions, the damage is not as extensive as with the recoil from fission products when the fuel is inside the reactor; the alpha particles will result primarily in displacements and interstitials associated with the oxygen sublattice. Additional inhomogeneity is caused by the implantation of He in the lattice. To simulate the effect of fission-product damage, requires the use of heavier ions, such as Zn, Mo, Cd, Sn, Xe, I, Pb, Au and U \cite{Matzke1996, Matzke2000}. This suggests dpa is not the only variable that should be considered. On the other hand, as Fig. \ref{fig:fig7} shows, we do observe a substantial increase in the phonon linewidths with the He$^{2+}$ irradiation we have performed, even of the acoustic modes. Presumably, the effects on the optic modes would be even greater if the main accommodation of damage is in the oxygen sublattice. Future experiments should look to observe the LO$_1$ mode, as it carries a fair proportion of the heat; however, measuring such optic modes represents a significant experimental challenge.

Observed swelling during this radiation damage experiment in the growth direction of $\sim$\,0.6\,\% with a dpa of 0.15, is in good agreement with that produced in the top layer of bulk UO$_2$ by Debelle \textit{et al}.,\cite{Debelle2011}, where they used 20\,keV alpha particles, i.e. an energy 100 times less than used in our experiments. With lower-energy particles, more He atoms will be implanted in the first few microns of the bulk UO$_2$ sample. Debelle \textit{et al}. show (Fig. 2 of Ref. \cite{Debelle2011}) that when the damage is increased to 3.3\,dpa the lattice loses definition and an average lattice swelling cannot be determined. At this high level of irradiation, small grain growth, and polygonization can be induced.  This is associated with the `high burn-up' structure, in which the grain sizes are reduced from microns to 100's of nm \cite{Rondinella2010}. Previous work suggest that the lattice becomes smaller when the `high burn-up' structure appears \cite{Spino2000}. At this stage of damage, where the microstructure has been significantly changed, the measurement of phonons by IXS would not be possible.

Further supporting evidence that thermal conductivity should change in our films is provided by the study of Weisensee \textit{et al}. \cite{Weisensee2013} in which 360\,nm epitaxial films of UO$_2$ were irradiated with 2\,MeV\,Ar$^+$ ions at room temperature, and the thermal conductivity decrease (of about $\sim$\,50\%) was measured directly with a time-domain thermal reflectance technique. These UO$_2$ films were grown on YSZ substrates \cite{Strehle2012}, where it is known that the UO$_2$ is under 6.4\,\% compressive strain. The authors of Ref. \cite{Weisensee2013} do not report a lattice swelling, but from their Fig. 1 this may be estimated at no more than $\sim$\,0.28\,\%. Of course, the growth direction should not be directly affected by the lattice mismatch, but this aspect does cast doubt as to whether this swelling is a meaningful measure, as indeed the authors themselves note. Weisensee \textit{et al}. have shown that with an irradiation dose of 10$^{15}$\,Ar$^+$\,/cm$^2$ the thermal conductivity drops by a factor of $\sim$\,2.5 and this does not change for a further increase of the dose by a factor of 10 [See Fig. 4 of Ref \cite{Weisensee2013}]. Fig. 4 shows that the decrease in thermal conductivity is already saturating by $\sim$\,10$^{15}$\,Ar$^+$\,/cm$^2$. The dose used in the current experiment is 6.7\,$\times$\,10$^{16}$ He$^{2+}$\,/cm$^2$ ions a factor of $\sim$\,7 times more than used by Weisensee \textit{et al}.  \cite{Weisensee2013} with the ions used (He$^{2+}$) lighter (by a factor of 10) than Ar$^+$. A direct relationship between these two experiments is difficult to quantitatively establish, although qualitatively the comparison with Ref.  \cite{Weisensee2013} suggests the thermal conductivity of our sample should drop by about 50\,\%.

In summary, as shown in Fig. 2 of Ref. \cite{Wiss2014}, our value of radiation damage (0.15\,dpa), coupled with a swelling of 0.6\,\%, does appear consistent with many current studies, and does suggest a reduction would be observed in the thermal conductivity. This agrees with the observations we have made in changes of the linewidths of the acoustic phonons.

Future irradiation to cause displacements in the uranium sublattice using heavier particles Xe for example \cite{Matzke2000, Usov2013, Popel2016}, to simulate what happens in actual irradiated fuels due to the fission recoil damage might show interesting changes to the phonon spectra.

\subsection{Phonons and their linewidths}
These experiments have been able to measure the acoustic phonons from a 300\,nm epitaxial film of UO$_2$. The grazing incidence technique has been refined so that the penetration depth into the sample is $\sim$ 150\,nm. This is a small volume of homogeneous damage, that may be roughly estimated as related to the cross section of the beam, multiplied by the attenuation length of 10 $\mu$m. This gives a mass of UO$_2$ (density of $\sim$ 10 gcm$^{-3}$) of $\sim$ 100 ng. For an inelastic neutron experiment samples would have to be at least 50\,mm$^3$, i.e. $\sim$\,0.5\,g, this gives an enormous increase in sensitivity for the X-ray experiments compared to neutron inelastic scattering. The intensity is increased by the large photon cross section from the 92 electrons around the U nucleus as well as by the greater X-ray flux. The optic modes, which primarily consist of motions of the oxygen atoms \cite{Pang2014, Maldonado2016}, could not be observed with IXS. This has been shown to be a limitation of the technique, as in the study of NpO$_2$ from a larger bulk sample in the X-ray beam \cite{Maldonado2016}.

To our knowledge there have been only two studies published using inelastic scattering to address the phonons of surfaces or in thin films.
(1) The work on NbSe$_2$ where the authors \cite{Murphy2005} used grazing-incident X-ray scattering with the angle of incidence set either below or above the critical angle to observe the soft-mode associated with the charge-density wave in this material. Their interest was primarily on the energy of these soft modes, and in a discussion of complicated inelastic spectra they make no comments on the phonon linewidths, although the probing distance for settings below the critical angle was only $\sim$ 4\,nm.
(2) Experiments on InN films \cite{Serrano2011} were performed also in grazing-incidence geometry, but a film of thickness 6.2 $\mu$m was used, some 20 times thicker than the films we have used and the penetration length of the X-rays was $\sim$\,50\,$\mu$m, so 5 times more than in our case. These larger parameters may be the reason they observe resolution-limited phonon linewidths.

As shown in Fig. \ref{fig:fig7} and Table A1, measured linewidths at 295\,K in the pristine sample of a 300\,nm UO$_2$ film are significantly larger (especially at higher energies) than reported at 295\,K for the bulk by Pang \textit{et al}. \cite{Pang2013, Pang2014}. We can be more certain of this increase over the value from bulk samples, as experiments have recently been performed \cite{Paolasini} on a small (bulk) single crystal of UO$_2$ on the same instrument (ID28), and with the same experimental set-up giving the resolution of 3\,meV. The linewidths (shown in Fig.\,\ref{fig:fig7}) deduced from these experiments \cite{Paolasini} are in good agreement (or even smaller) with those deduced by Pang \textit{et al}. \cite{Pang2013, Pang2014}. Initially, it may be thought that these differences can be attributed to finite size effects, however this seems unlikely given that the penetration distance is 150\,nm, and the chemical unit cell is 0.547\,nm. We suggest that this difference is due to intrinsic strain in the pristine film causing a decrease in the phonon lifetime when the phonon wavelengths become short, i.e. at higher energies near the zone boundaries. This accounts for the slope of the linewidth vs energy curve (Fig. \ref{fig:fig7}) for the film data. The effect is strongly enhanced when the irradiated film is considered, due to greater strain and the presence of inhomogeneity caused by the He particles in the lattice. This aspect, as well as the changes in lifetime in the phonons due to near-surface effects would be interesting to consider theoretically. Measurements on unirradiated bulk samples \cite{Pang2013, Pang2014} show a slope in the linewidth vs energy curves for most phonon branches (see Fig.\,2 of Ref. \cite{Pang2013} and Fig.\,7 of Ref. \cite{Pang2014}). These experiments on irradiated film (as shown in our Fig.\,\ref{fig:fig7} and Table A1) suggest the slopes for irradiated samples may be greater than with the sample at higher temperature, and it would be interesting to explore theoretically the effects of defects on the linewidth vs energy and momentum dependence.

As discussed in Ref. \cite{Pang2013, Pang2014} the thermal conductivity $\kappa_{\textbf{q}j}$ for phonons of wave vector \textbf{q} and branch j is given by

\begin{equation}
\kappa_{\textbf{q}j} = (1/3)C_{\textbf{q}j}v^2_{\textbf{q}j}/\Gamma_{\textbf{q}j}
\end{equation}

where C$_{\textbf{q}j}$ is the phonon heat capacity and v$_{\textbf{q}j}$ = $\delta$E$_{\textbf{q}j}$/$\delta$\textbf{q} is the group velocity determined by the local dispersion gradients. The phonon mean free paths $\lambda_{\textbf{q}j}$ = v$_{\textbf{q}j}\tau_{\textbf{q}j}$ depend on the measured phonon linewidths through the relaxation time $\tau_{\textbf{q}j}$ = 1$/\Gamma_{\textbf{q}j}$.

In our case, since the energies of the phonons have not changed between the pristine and irradiated samples, the change in thermal conductivity will depend only on the change in the linewidths. A complete calculation of the thermal conductivity of the damaged films is not possible without a measure of the optic phonon linewidth. However, given the acoustic modes contribute $\sim$\,50\,\% to the thermal conductivity \cite{Pang2013}, therefore the doubling of the average linewidths (see Fig. \ref{fig:fig7}) translates to a factor of two drop in the thermal conductivity for damaged films. This is consistent with that found by Weisensee \textit{et al.}\cite{Weisensee2013} for similar films irradiated by Ar ions.

\subsection{Alternative methods for phonon measurements}
Raman scattering is an alternative method used to obtain phonon measurements, and such studies have observed the LO modes from a number of different samples \cite{Livneh2006, Livneh2008}. If we confine our attention to the low-energy modes, then Ref. \cite{Desgranges2012, Manara2003} show that the only really strong line is that at 445\,cm$^{-1}$ = 55.2\,meV, which corresponds to the TO$_2$ phonon line at the zone-center ($\Gamma$) in Fig.\,\ref{fig:fig4}. As shown by Pang \textit{et al.} \cite{Pang2013} (see Fig. 4 of Ref. \cite{Pang2013}) this TO$_2$ mode contributes very little to the thermal conductivity. Therefore Raman technique will not elucidate the role of damage in changes of thermal conductivity. Due to the penetration depth of the laser light Raman studies performed on irradiated UO$_2$ \cite{Manara2003, Guimbretiere2012,Guimbretiere2013,Desgranges2014} samples, require $\sim$\,100\,$\mu$m of UO$_2$, far more than the 300\,nm film used in this experiment. Any backscattering Raman technique on a 300 nm thin film will be sensitive to the substrate excitations, so a grazing-incidence geometry would be required.

Desgranges and colleagues have irradiated UO$_2$ samples with high energy ($>$
\,20\,MeV) He$^{2+}$ ions and are able \textit{in situ} to probe the Raman signal with a spatial resolution of about 2\,$\mu$m over the irradiated depth profile of 150\,$\mu$m. They observe extra peaks in the Raman signal, which they associate \cite{Desgranges2014} with defects due to a charge-separated state, where U$^{3+}$ and U$^{5+}$ ions coexist in the irradiated material. They also observe \cite{Guimbretiere2012} the TO$_2$ phonon (T$_{2g}$ mode) dropping in frequency near the Bragg peak of the damage at 130\,$\mu$m, but this drop in energy is only 0.5\,cm$^{-1}$, i.e. $<$\,0.1\,meV, which is far smaller than our resolution with IXS of 3\,meV. Comparison with these Raman papers \cite{Guimbretiere2012, Guimbretiere2013} is difficult due to the lack of other characterisation information such as a value for the lattice swelling, $\Delta$a/a.

\section{Conclusions}
These experiments shed further light on the reasons for the drop of the thermal conductivity in UO$_2$ when it is irradiated in a reactor, which is a technical problem when using UO$_2$ fuel. We have shown that irradiation with He$^{2+}$ ions on a thin epitaxial film produces uniform damage (Fig. \ref{fig:fig2}) over the whole film thickness, and characterisation of the (homogeneous) damage is equivalent to 0.15\,dpa, with a swelling of the damaged UO$_2$ by $\Delta$a/a $\sim$ 0.6\,\%, (Fig. \ref{fig:fig3}), i.e. a volume swelling of $\sim$\,2\,\%. (This assumes that the clamping by the SrTiO$_3$ substrate allows the film to expand equally in all three directions.)  There will also be inhomogeneous damage caused by the presence of He particles in the lattice, as well as displaced oxygen atoms.

We have succeeded in measuring the acoustic phonons (Figs. \ref{fig:fig5} and \ref{fig:fig6}) by grazing-incidence X-ray scattering from thin films, where the estimated amount of material giving the phonons is $\sim$ 100 ng. The optic phonons, some of which are known to be important in carrying the heat in UO$_2$ \cite{Pang2013, Pang2014}, could not be measured as they are about 100 or more times weaker than the acoustic modes \cite{Maldonado2016}. The acoustic modes, both transverse and longitudinal, were measured with enough precision to analyse their respective widths (Fig. \ref{fig:fig7} and Table A1).

For both the pristine and irradiated films (Fig. \ref{fig:fig6}), the energies of the acoustic phonons are within experimental error consistent with those of the bulk. This is not surprising given that the UO$_2$ phonon spectra changes only a small amount when heating from 295 to 1200 K \cite{Pang2013, Pang2014}, in which case the volume expansion is comparable to that caused by the damage induced in this study through He$^{2+}$ irradiation.

A definite increase in the phonon lifetimes is observed for phonons in the pristine sample as compared to the bulk values, as measured by both neutron and X-ray inelastic scattering. We attribute these changes to strain in the pristine films.

Changes in the linewidths between the pristine and damaged films are shown in a plot of deconvoluted FWHM vs energy for the TA(100) modes, see Fig. \ref{fig:fig7} and Table A1. All acoustic modes show significant effects, with the average effect being an increase in FWHM of about 50-100\%, depending on the energy. As phonon energies do not change, and the group velocity of the phonons is the same for the pristine and damaged samples, this can be translated directly into a decrease in the contribution to the thermal conductivity for the low-energy acoustic modes for the damaged UO$_2$ thin film \cite{Pang2013, Weisensee2013}. Total thermal conductivity cannot be deduced from these experiments without measuring the higher-energy optic modes, and especially the LO$_1$ mode that carries so much of the heat \cite{Pang2013} in UO$_2$. The measurement of the acoustic modes suggest a significant decrease in thermal conductivity of irradiated UO$_2$ is caused by the damage affecting the lifetime of the phonons, and not by other possible mechanisms due to increased grain boundaries and defects.

We hope this work prompts more careful theoretical analysis of the thermal conductivity of UO$_2$ in the future, as well as further experiments as the intensity of synchrotron X-ray sources increases. It will be interesting, for example to irradiate films with heavier ions to see the additional effects on the phonon spectra.

To develop grazing incidence Raman scattering capable of examining irradiated films of $<$\,1\,$\mu$m would also lead to further progress, and would be a great help to monitor films already damaged.
\begin{table}[]
\centering
\caption{Results of analysis of the energies and full-width at half maximum (here given as 2$\Gamma$ of the DHO function used for the fitting, similarly to Ref. \cite{Pang2014}) of the acoustic phonons in the pristine and irradiated films. Notice that because of the 0.5$^{\circ}$ tilt of the film, the value of the L component is not strictly zero, so is indicated here as $\delta$, where this varies between 0.05 to 0.15 depending on the phonon.}
\label{my-label}
\begin{tabular}{p{1.6cm} p{1.6cm} p{1.6cm} p{1.6cm} p{1.6cm} p{1.6cm}}
\hline\hline
\multicolumn{5}{c}{TA}                                                                                                                                                                                                                                                                                                             \\
\hline
\begin{tabular}[c]{@{}c@{}}Wave\\   vector\end{tabular}  & \begin{tabular}[c]{@{}c@{}}E$_{PRIS}$\\   (meV)\end{tabular} & \begin{tabular}[c]{@{}c@{}}2$\Gamma_{PRIS}$\\   (meV)\end{tabular} & \begin{tabular}[c]{@{}c@{}}E$_{IRRAD}$\\   (meV)\end{tabular} & \begin{tabular}[c]{@{}c@{}}2$\Gamma_{IRRAD}$\\   (meV)\end{tabular} \\
\hline
(4 0.2 $\delta$)                                         & 5.0(2)                                                       &    -                                                                  & 3.7(2)                                                        &    -                                                                   \\
(4 0.4 $\delta$)                                         & 6.5(1)                                                       & 0.9(1)                                                             & 6.9(1)                                                        & 1.3(2)                                                              \\
(4 0.5 $\delta$)                                         &    -                                                            &    -                                                                  & 8.6(1)                                                        & 1.6(2)                                                              \\
(4 0.6 $\delta$)                                         & 11.0(1)                                                      & 1.3(2)                                                             & 10.1(1)                                                       & 2.2(2)                                                              \\
(4 0.8 $\delta$)                                         & 13.2(1)                                                      & 1.7(4)                                                             & 12.5(1)                                                       & 3.0(4)                                                              \\
(4 1.0 $\delta$)                                         & 14.0(1)                                                      & 1.5(2)                                                             & 13.1(1)                                                       & 3.1(4)                                                              \\
                                                         &                                                              &                                                                    &                                                               &                                                                     \\
(2.2 1.8 $\delta$)                                       & 6.7(1)                                                       &    -                                                                  & 6.5(1)                                                        &    -                                                                   \\
(2.4 1.6 $\delta$)                                       & 10.3(1)                                                      & 0.8(1)                                                             & 10.1(1)                                                       & 1.0(3)                                                              \\
(2.6 1.4 $\delta$)                                       & 11.9(1)                                                      & 1.0(1)                                                             &    -                                                             &    -                                                                   \\
(2.8 1.2 $\delta$)                                       & 12.6(1)                                                      & 1.5(1)                                                             & 12.9(1)                                                       & 1.9(4)                                                              \\
(3.0 1.0 $\delta$)                                       & 12.7(1)                                                      & 1.1(1)                                                             & 13.3(2)                                                       & 2.8(5)                                                              \\
                                                         &                                                              &                                                                    &                                                               &                                                                     \\
\hline\hline
\multicolumn{5}{c}{LA}                                                                                                                                                                                                                                                                                                             \\
\hline
\begin{tabular}[c]{@{}c@{}}Wave\\   vectors\end{tabular} & \begin{tabular}[c]{@{}c@{}}E$_{PRIS}$\\   (meV)\end{tabular} & \begin{tabular}[c]{@{}c@{}}2$\Gamma_{PRIS}$\\   (meV)\end{tabular} & \begin{tabular}[c]{@{}c@{}}E$_{IRRAD}$\\   (meV)\end{tabular} & \begin{tabular}[c]{@{}c@{}}2$\Gamma_{IRRAD}$\\   (meV)\end{tabular} \\
\hline
(4.2 0 $\delta$)                                         & 9.7(2)                                                       & -                                                                  & 8.7(1)                                                        & -                                                                   \\
(4.4 0 $\delta$)                                         & 15.6(1)                                                      & 1.1(2)                                                             & 15.8(1)                                                       & 1.9(4)                                                              \\
(4.6 0 $\delta$)                                         & 20.4(1)                                                      & 1.3(2)                                                             & 20.9(2)                                                       & 1.9(4)                                                              \\
(4.8 0 $\delta$)                                         & 23.3(1)                                                      & 2.0(3)                                                             & 24.0(2)                                                       & 3.0(6)                                                              \\
(5.0 0 $\delta$)                                         & 24.3(1)                                                      & 2.5(3)                                                             & 25.0(2)                                                       & 3.4(8)                                                              \\
                                                         &                                                              &                                                                    &                                                               &                                                                     \\
(2.2 2.2 $\delta$)                                       & 11.5(1)                                                      & -                                                                  & 11.5(2)                                                       & 2.8(8)                                                              \\
(2.4 2.4 $\delta$)                                       & 18.0(2)                                                      & 0.7(2)                                                             & 17.1(6)                                                       & 3.1(8)                                                              \\
(2.6 2.6 $\delta$)                                       & 20.1(1)                                                      & 1.0(4)                                                             & 20.0(6)                                                       & -                                                                   \\
(2.8 2.8 $\delta$)                                       & 15.7(1)                                                      & 1.1(3)                                                             & 15.4(2)                                                       & 2.2(6)                                                              \\
(3.0 3.0 $\delta$)                                       & 12.9(2)                                                      & 1.3(4)                                                             & 13.4(2)                                                       & 2.4(3)

\end{tabular}
\end{table}

\section{Acknowledgements}

We thank Simon Pimplott and Tom Scott for their support and encouragement. SR would like to thank the AWE and EPSRC for funding. GHL thanks Vincenzo Rondinella, Thierry Weiss, Roberto Cacuiffo, and Nicola Magnani at JRC, Karlsruhe for discussions about the damage in UO$_2$ and the response of the phonons. Discussions with Boris Dorado and Johann Bouchet of the CEA on the effects of defects on the phonon dispersion relations are much appreciated.

\bibliography{Phonon_Paper2}

\end{document}